\documentclass[prb,twocolumn, flushbottom, superscriptaddress]{revtex4-1}
\usepackage{amsmath,amssymb}
\usepackage{graphicx}
\usepackage{color}
\usepackage{times}
\usepackage[colorlinks,bookmarks=false,citecolor=blue,linkcolor=red,
urlcolor=blue]{hyperref}

\usepackage{float}

\begin{document}

\title{Phase diagram of the $J_{1}$-$J_{2}$ Heisenberg model on 
the kagome lattice}

\author{F.~Kolley}
\affiliation{Department of Physics and Arnold Sommerfeld Center 
for Theoretical Physics, Ludwig-Maximilians-Universit\"at M\"unchen, 
D-80333 M\"unchen, Germany}
\author{S.~Depenbrock}
\affiliation{Department of Physics and Astronomy, University of 
California, Irvine, CA 92697, USA}
\author{I.~P.~McCulloch}
\affiliation{School of Physical Sciences, The University of Queensland, 
Brisbane, QLD 4072, Australia}
\author{U.~Schollw\"ock}
\affiliation{Department of Physics and Arnold Sommerfeld Center for 
Theoretical Physics, Ludwig-Maximilians-Universit\"at M\"unchen, D-80333 
M\"unchen, Germany}
\author{V.~Alba}
\affiliation{Department of Physics and Arnold Sommerfeld Center 
for Theoretical Physics, Ludwig-Maximilians-Universit\"at M\"unchen, 
D-80333 M\"unchen, Germany}

\date{\today}

\begin{abstract}
We perform an extensive density matrix renormalization group (DMRG) study of the 
ground-state phase diagram of the spin-$1/2$ $J_{1}$-$J_{2}$ Heisenberg model on 
the kagome lattice. We focus on the region of the phase diagram around 
the kagome Heisenberg antiferromagnet, i.e., at $J_2=0$. We investigate the static 
spin structure factor, the magnetic correlation lengths, and the spin gaps. 
Our results are consistent with the absence of magnetic order in a narrow region 
around $J_2\approx 0$, although strong finite-size effects  do not allow us to  
accurately determine the phase boundaries. This result is 
in agreement with the presence of an extended spin-liquid region, 
as it has been proposed recently. Outside the disordered region, we find that  
for ferromagnetic and antiferromagnetic $J_2$ the ground state displays signatures 
of the magnetic order of the $\sqrt{3}\times\sqrt{3}$ and the $q=0$ type, 
respectively. Finally, we focus on the structure of the entanglement spectrum (ES) 
in the $q=0$ ordered phase. We discuss the importance of the choice of the bipartition 
on the finite-size structure of the ES. 
\end{abstract}

\maketitle

\section{Introduction}
\label{sec1}

The nature of the ground state of the antiferromagnetic spin-$1/2$ Heisenberg model 
on the kagome lattice (KHA) has been debated for a long time. Despite substantial 
analytical and numerical effort no agreement has been reached yet in the community. 
The proposed ground states include several valence bond crystals (VBC)~\cite{Zeng1990,
Marston1991, Hastings2000, Nikolic2003, Singh2007, Singh2008, Iqbal2012}, and both 
gapped and gapless spin liquids \cite{Kalmeyer1989, Sachdev1992, Yang1993, Misguich2002, 
Wang2006, Ryu2007, Ran2007, Hermele2008, Jiang2008, Iqbal2011, Lu2011, Huh2011, 
Messio2012, He2014b, He2014c, Gong2014a, Zhu2014, Bauer2014}.

However, recent DMRG simulations~\cite{White1992,Schollwock2005,Schollwock2011} provided 
convincing evidence that the ground state of the KHA is a gapped spin liquid with topological 
entanglement entropy $\gamma=\log(2)$~\cite{Yan2011,Depenbrock2012,Jiang2012}. This is 
compatible with both a spin liquid of the toric-code or the double-semion~\cite{Freedman2004,
Levin2005} type. Although the former appears naturally in mean field theories 
of the KHA~\cite{Sachdev1992}, and for quantum dimer models on the kagome lattice~\cite{Poilblanc2012, 
Schuch2012} and was therefore favored, recent numerical studies provide indirect evidence that the ground 
state of the KHA is in a double-semion phase~\cite{He2014a,Qi2014,Buerschaper2014,
Iqbal_M2014}. This was motivated by the observation of a chiral spin liquid phase adjacent to 
the $Z_2$ phase~\cite{He2014a,Gong2014a}. Notice, however, that a recent theoretical analysis 
rules out the double semion scenario~\cite{Zalatel2014}. 

\begin{figure}[t]
\includegraphics[width=.9\linewidth]{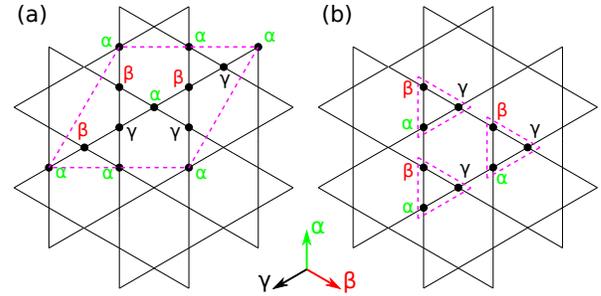}
\caption{
 (Color online) Ordering patterns of the classical $J_1$-$J_2$ Heisenberg model 
 on the kagome lattice ($J_1$-$J_2$ KHA). The orientations of the spins of the three 
 ferromagnetic sublattices are denoted as $\alpha$, $\beta$ and $\gamma$. Spins in 
 different sublattices form an angle of $2\pi/3$. (a) The $\sqrt{3}\times\sqrt{3}$ 
 state arising at $J_{2}\ll 0$. (b) The $q=0$ state, which appears for $J_{2}\gg 0$. 
 The dashed lines highlight the unit cells. 
}
\label{fig:classical-order}
\end{figure}

Moreover, it has been suggested that the spin liquid behavior survives upon introducing 
a small antiferromagnetic next-nearest-neighbor interaction~\cite{Jiang2012}, i.e., in the 
$J_1$-$J_2$ Heisenberg model ($J_1$-$J_2$ KHA). This is in contrast with the $T=0$ phase 
diagram of the {\it classical} version of the model. At $J_2=0$ the ground state of the 
classical $J_1$-$J_2$ KHA exhibits an extensive degeneracy~\cite{Chalker1992,Huse1992,
Ritchey1993, Messio2011, Cepas2011, Spenke2012, Chern2013}. This is lifted upon introducing an 
infinitesimal $J_2$, and the system develops magnetic order~\cite{Harris1992}. Precisely, for 
ferromagnetic $J_{2}$ the so-called $\sqrt{3}\times\sqrt{3}$ order emerges, whereas in the 
antiferromagnetic case one has the $q=0$ order. The two ordering patterns are shown schematically 
in Figure~\ref{fig:classical-order}. 
The magnetic order survives in the quantum model, at least for large enough $J_2$, as it has 
been established by exact diagonalization studies~\cite{Lecheminant1997}. 
However, the precise phase boundary between the magnetically ordered phases and the disordered 
spin-liquid region at $J_2\approx 0$ has not been determined yet (see Ref.~\onlinecite{Suttner2014} 
for some interesting results obtained using the functional renormalization group approach).
%
%

In this work by performing $SU(2)$-symmetric DMRG calculations we investigate the 
ground-state phase diagram of the $J_1$-$J_2$ KHA as a function of $J_1$ and $J_2$. Here we set 
$J_1=1$, considering both positive and negative $J_2$. We study the finite-size behavior 
of the static spin structure factor, the spin-spin correlation length, and the spin gap. For 
ferromagnetic $J_2$ we provide numerical evidence that magnetic  order of the $\sqrt{3}\times\sqrt{3}$ 
type survives up to $J_2\lesssim -0.1$. On the other hand, for antiferromagnetic $J_2$ signatures 
of the $q=0$ state appear already at $J_2\gtrsim 0.2$. In the narrow region at $-0.1\lesssim J_2
\lesssim 0.2$, although strong finite-size effects are present, our data are compatible with an 
extended disordered region, suggestive of a spin liquid behavior~\cite{Jiang2012}. Finally, we analyze 
the structure of the entanglement spectrum (ES)~\cite{Li2008} in the $q=0$ ordered phase at large 
$J_2\gg 0.2$. Recently, it has been suggested that in presence of continuous symmetry breaking the low-lying 
levels in the ES are reminiscent of the so-called tower-of-states, which appear in finite-size
{\it energy} spectra~\cite{Metlitski2011,Alba2013,Kolley2013}. This correspondence has been  
checked numerically in Ref.~\onlinecite{Kolley2013} for the $J_1$-$J_2$ KHA at large ferromagnetic $J_2$, 
i.e., in presence of the $\sqrt{3}\times\sqrt{3}$ order, and for the $2D$ Bose Hubbard model in the 
superfluid phase~\cite{Alba2013}. Here we investigate how the identification of the correct 
tower-of-states structure in the ES depends on the choice of the bipartition, in finite-size systems. 

The article is organized as follows. Section \ref{sec2} introduces the $J_{1}$-$J_{2}$ Heisenberg 
model on the kagome lattice, and the DMRG method. In particular, we describe in detail the geometry 
used in the DMRG simulations. In section~\ref{sec3} and~\ref{sec4} we discuss the numerical results 
for the static spin structure factor, and the spin-spin correlation length. The energy gaps are 
presented in section~\ref{sec5}. Finally, in section~\ref{sec7} we investigate the structure of 
the entanglement spectrum in the $q=0$ ordered phase. 

\begin{figure}[t]
\includegraphics[width=.9\linewidth]{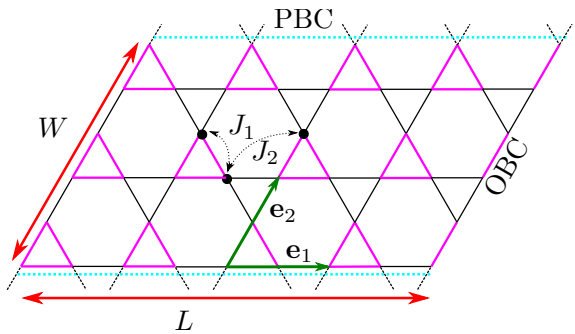}
\caption{
 (Color online) The $J_{1}$-$J_{2}$ Heisenberg model on kagome 
 cylinders. The spins are located at the vertices of the lattice. 
 The two basis vectors of the lattice are denoted as $\mathbf{e}_{1}$ 
 and $\mathbf{e}_{2}$. Periodic (PBC) and open (OBC) boundary conditions 
 are imposed along the $\mathbf{e}_{2}$ and $\mathbf{e}_{1}$ directions, 
 respectively. The unit cells consist of three sites and are denoted by 
 the thicker (purple) triangles. $J_{1}$ and $J_2$ (see arrows) are the 
 interaction strengths between nearest and next-nearest-neighbor 
 sites, respectively. The figure shows a cylinder with width $W=3$ 
 (YC6 geometry) and length $L=4$. Here $W$ and $L$ denote the number of unit cells 
 in the $\mathbf{e}_2$ and $\mathbf{e}_1$ directions, respectively. Notice 
 that the unit cells at the right boundary are incomplete in order to 
 alleviate edge effects in the DMRG simulation.
}
\label{fig:lattice}
\end{figure}

\section{Model and Method}
\label{sec2}

The spin-$1/2$ $J_1$-$J_2$ Heisenberg model on the kagome lattice is 
defined by the  SU(2)-invariant Hamiltonian 
\begin{align}
\label{ham}
\mathcal{H} = J_{1} \sum_{\langle i,j \rangle} \mathbf{S}_{i} 
\cdot \mathbf{S}_{j} + J_{2} \sum_{\langle \langle i,k \rangle 
\rangle} \mathbf{S}_{i} \cdot \mathbf{S}_{k}.
\end{align}
Here, $\mathbf{S}_{i}$ is the spin operator acting on the lattice site $i$, while $\langle i,j
\rangle$ and $\langle \langle i, k \rangle \rangle$ denote nearest and next-nearest-neighbor 
sites, respectively. We restrict ourselves to $J_{1}=1$ in~\eqref{ham}.

We obtain the ground state of the $J_1$-$J_2$ KHA using $SU(2)$-symmetric DMRG calculations. 
The geometry used in the simulations is depicted in Figure~\ref{fig:lattice}. The two basis 
vectors of the kagome lattice are denoted as $\mathbf{e}_1,\mathbf{e_2}$. The unit cell 
(thicker purple lines) contains three sites. Since DMRG prefers open boundary conditions, 
we consider kagome cylinders, using periodic (open) boundary conditions in the 
$\mathbf{e}_{2}$ ($\mathbf{e}_1$)-direction. Here we focus on cylinders with  width $W$ 
and length $L$, where $W$ and $L$ are the numbers of unit cells along the $\mathbf{e}_{2}$ 
and $\mathbf{e}_{1}$ directions, respectively. In order to alleviate spurious effects due to 
sharp edges, the unit cells at the right boundary of the cylinder contain only two sites. In the Appendix we show that the results are qualitatively the same for lattices with integer number of unit cells and fully periodic tori.
The total number of spins on the lattice used for the main text is given as $W\times (3L+2)$. Here we consider only 
cylinders with $W=3$ and $W=4$, which, following Ref.~\onlinecite{Yan2011}, are referred 
to as YC6 and YC8 cylinders. The computational time scales approximately linearly 
with $L$ and exponentially with $W$. In our DMRG calculations we keep up to $\sim 5000$ 
$SU(2)$ states, which correspond to approximately 20000 $U(1)$ states. This allows us to obtain accurate 
ground-state wavefunctions for cylinders with lengths $L=4,6,8,10,12$ for both the YC6 and 
YC8 geometries. The largest cylinder considered in this work (with $W=4$ and $L=12$) contains 
$152$ spins.

\section{Static spin structure factor}
\label{sec3}

Here we discuss the static spin structure factor $S(\mathbf{q})$ obtained from 
the ground state of the $J_1$-$J_2$ Heisenberg model as a function of $-0.2
\leq J_2\leq 0.4$. The structure factor is defined as 
\begin{align}
\label{sq} 
S(\mathbf{q}) = \frac{1}{N}\sum_{i,j=1}^{N} \langle \mathbf{S}_{i}
\cdot \mathbf{S}_{j} \rangle e^{i\mathbf{q}\cdot (\mathbf{r}_{i}-
\mathbf{r}_{j})}.
\end{align}
Here $N$ is the total number of lattice sites, $\langle \cdot \rangle$ denotes   
the ground-state expectation value, $\mathbf{r}_{i}$ is the position of site $i$, 
and $\mathbf{q}$ is a generic vector in the reciprocal lattice.

\begin{center}
\begin{figure*}[tb]
\includegraphics[width=1.\linewidth]{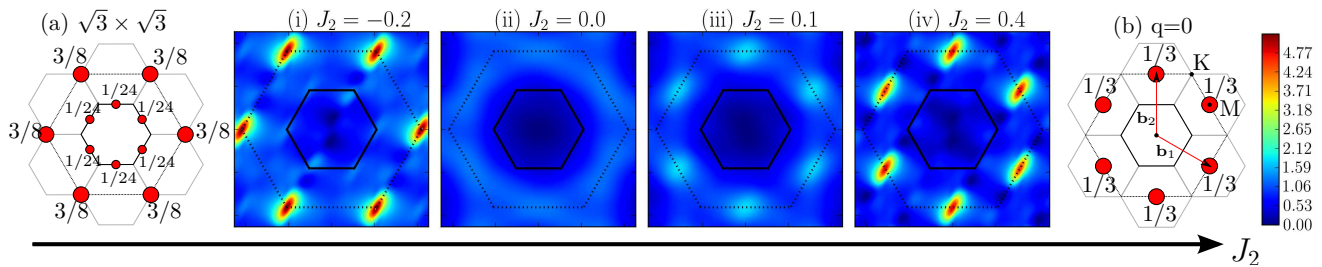}
\caption{(Color online) The static spin structure factor 
 $S(\mathbf{q})$ obtained from ground-state DMRG simulations 
 of the $J_{1}$-$J_{2}$ Heisenberg model on kagome cylinders 
 with $3\times 12$ unit cells (YC6) and several values of $J_{2}$: 
 (i) $J_{2}=-0.2$, (ii) $J_{2}=0.0$, (iii) $J_{2}=0.1$, (iv) 
 $J_{2}=0.4$. The solid and the dotted lines show the first and the extended 
 Brillouin zones, respectively. (a) and (b) show the expected 
 structure factors for the {\it classical} $\sqrt{3}\times\sqrt{3}$ 
 and $q=0$ states, respectively. In (b) $\mathbf{b}_{1}$ and 
 $\mathbf{b}_{2}$ form a basis for the reciprocal lattice, while
 $K$ and $M$ are the high-symmetry points. The circles denote the 
 peaks in the structure factors, whereas the numbers are the 
 relative peak heights. Clearly, DMRG data at $J_2=-0.2$ and 
 $J_2=0.4$ match the expected structure factors for the $\sqrt{3}
 \times\sqrt{3}$ and the $q=0$ states. On the  other hand at 
 $J_2\approx 0$ the height of the peaks in the structure factor 
 is vanishing (see (ii)), which is compatible with the absence of 
 magnetic order.
}
\label{fig:structurefactor}
\end{figure*}
\end{center}

Figure~\ref{fig:structurefactor} (a) and (b) show the expected structure 
factors (the circles denote the positions of the peaks in momentum space) for 
the classical $\sqrt{3}\times\sqrt{3}$ state and the $q=0$ state, respectively. Panels (i)-(iv) 
plot the DMRG result for $S(\mathbf{q})$ for $J_{2}=-0.2$, $J_2=0.0$, $J_2=0.1$ and $J_2=0.4$. 
The data are for a YC6 cylinder  (with $3\times 12$ unit cells, cf. Figure~\ref{fig:lattice}). 
Clearly, for $J_{2}=-0.2$ sharp peaks with $S(\mathbf{q}_K)\approx 6$ are visible at the 
$K$-points $\mathbf{q}_K$ of the extended Brillouin zone (see Figure~\ref{fig:structurefactor} 
(a), and Figure~\ref{fig:structurefactor} 
for the definition of the high-symmetry points), in agreement with what is expected for 
the $\sqrt{3}\times\sqrt{3}$ state. Notice that the 
much smaller peaks at the $M$-points of the first Brillouin zone cannot be resolved with 
the available system sizes. We observe that at $J_2=0$ $S(\mathbf{q})$ is featureless 
(see Figure~\ref{fig:structurefactor} (ii)), which signals the absence of magnetic order. On the 
other hand, already at $J_{2}=0.1$ some peaks start developing at the $M$-points $\mathbf{q}_M$, 
as expected for the classical $q=0$ state (cf. Figure~\ref{fig:structurefactor} 
(b)). These become sharper upon increasing $J_2$ (one has $S(\mathbf{q}_M)\approx 5$ for 
$J_{2}=0.4$).

All these features are more quantitatively discussed in Figure~\ref{fig:orderparameter} 
plotting the (squared) antiferromagnetic order parameter $m^2_{\mathbf{Q}}\equiv S(
\mathbf{Q})/N$ versus $J_2$. Here $\mathbf{Q}$ denotes the positions of the peaks of the 
structure factors. Data are for both YC6 and YC8 cylinders (panels (a) and (b) in the 
Figure) with lengths $L=4,6,8,10,12$. Precisely, Figure~\ref{fig:orderparameter} plots 
$m^2_{\mathbf{q}_K}$ for the $\sqrt{3}\times\sqrt{3}$ order for $J_2<0$ (empty symbols) and 
$m^2_{\mathbf{q}_M}$ (i.e., the order parameter for the $q=0$ order) for $J_2\ge0$. In the 
region $-0.1<J_2<0.2$, $m^2_\mathbf{Q}$ is almost featureless and $S(\mathbf{Q})$ itself is 
nearly size independent. This is compatible with a vanishing order parameter in the 
thermodynamic limit, as expected in a disordered phase. On the other hand, outside this region $S(\mathbf{Q})$ (and as a consequence $m^2_\mathbf{Q}$) exhibits a stronger dependence on the cylinder size.

A sharp increase of the order parameter can be observed for $J_2\approx -0.1$  
and $J_2\approx 0.2$, which could signal a phase transition in the thermodynamic limit. 
Surprisingly, while for $J_2\gtrsim 0.2$ $m^2_\mathbf{Q}$ increases with 
$W$, for $J_2\lesssim -0.1$ it slightly decreases. However, this could 
be attributed to strong finite-size corrections due to the fact that the YC8 
geometry is not commensurate with the large unit cell of the $\sqrt{3}\times
\sqrt{3}$ pattern (cf. Figure~\ref{fig:classical-order}). We anticipate that this 
change in the behavior of the order parameter at $J_2\approx-0.1$ and $J_2\approx 0.2$ 
is reflected in the triplet gap (cf. section~\ref{sec5}). In a magnetically 
ordered phase, for large system sizes one should expect 
$S(\mathbf{Q})/N=m^2_{\mathbf{Q},\infty}+a/\sqrt{N}+b/N+\dots$, with $m^2_{\mathbf{Q},\infty}$ 
the order parameter in the thermodynamic limit. Although a 
finite-size scaling analysis would allow to extract $m_{\mathbf{Q},\infty}$, 
providing  conclusive evidence for the presence of magnetic order at $J_2\ll -0.1$ and 
$J_2\gg 0.2$, it would require much larger system sizes than the ones currently available. 
Finally, from Figure~\ref{fig:orderparameter} one should observe that at fixed $W$, 
$m^2_\mathbf{Q}$ decreases with the cylinder length $L$, which might signal a   
vanishing order parameter in the limit $L\to\infty$, as expected, since infinitely long 
cylinders should exhibit $1D$ behavior. 

\section{Spin-spin correlation lengths}
\label{sec4}

From the structure factor $S(\mathbf{q})$ one can define a correlation 
length $\xi(\mathbf{Q},\mathbf{q}_{min})$ as~\cite{Sandvik2010, Pelissetto2002}
\begin{align}
\label{eq:corrlength}
\xi(\mathbf{Q},\mathbf{q}_{min}) =\frac{1}{|\mathbf{q}_{min}|}
\sqrt{\frac{S(\mathbf{Q})}{S(\mathbf{Q}+\mathbf{q}_{min})} - 1 },
\end{align}
where $\mathbf{q}_{min}$ is the point next to the peak (at $\mathbf{Q}$)  
of the structure factor. Here we choose $\mathbf{q}_{min}=\mathbf{b}_{1}/L$,  
with $\mathbf{b}_{1}$ being the reciprocal lattice vector corresponding to 
the long direction of the cylinder (see Figure~\ref{fig:structurefactor} (b) 
for its definition). Other choices of $\mathbf{q}_{min}$ are expected to be 
equivalent in the $2D$ limit $W,L\to\infty$. 

\begin{figure}[htb]
\includegraphics[width=.95\linewidth]{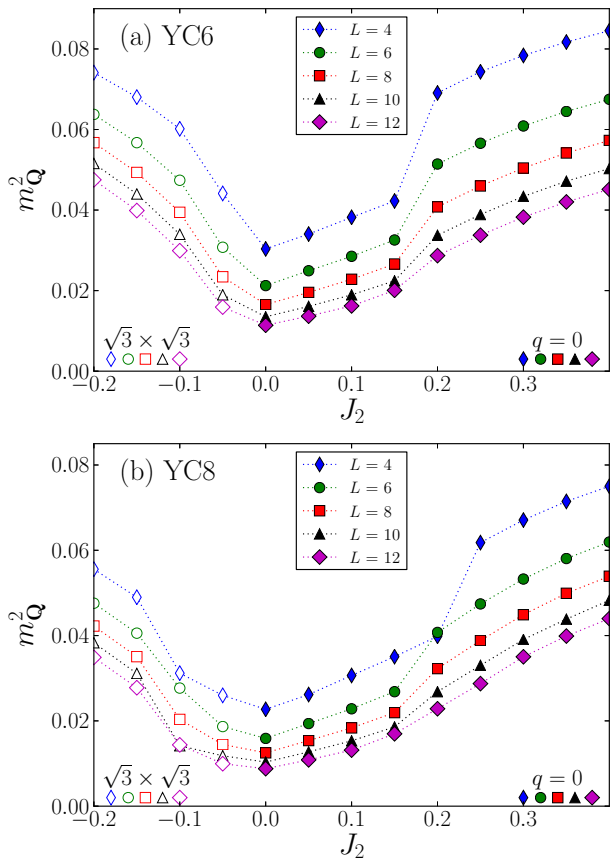} 
\caption{
 (Color online) The antiferromagnetic order parameter $m^{2}_\mathbf{Q}\equiv 
 S(\mathbf{Q})/N$ for the ground state of the $J_1$-$J_2$ Heisenberg model 
 plotted as a function of $J_{2}$. Here $\mathbf{Q}$ denotes the position of the 
 peak in the structure factor. Panels (a) and (b) correspond to the YC6 and YC8 
 cylinders, respectively. Empty symbols at $J_2<0$ are obtained from the 
 peak at the $K$-point in the extended Brillouin zone (corresponding to the 
 $\sqrt{3}\times\sqrt{3}$ state, see Figure~\ref{fig:structurefactor} (b)), 
 while full symbols at $J_2\gtrsim 0$ correspond to the $M$-point ($q=0$ state). 
}
\label{fig:orderparameter}
\end{figure}

\begin{figure}[t]
\includegraphics[width=.91\linewidth]{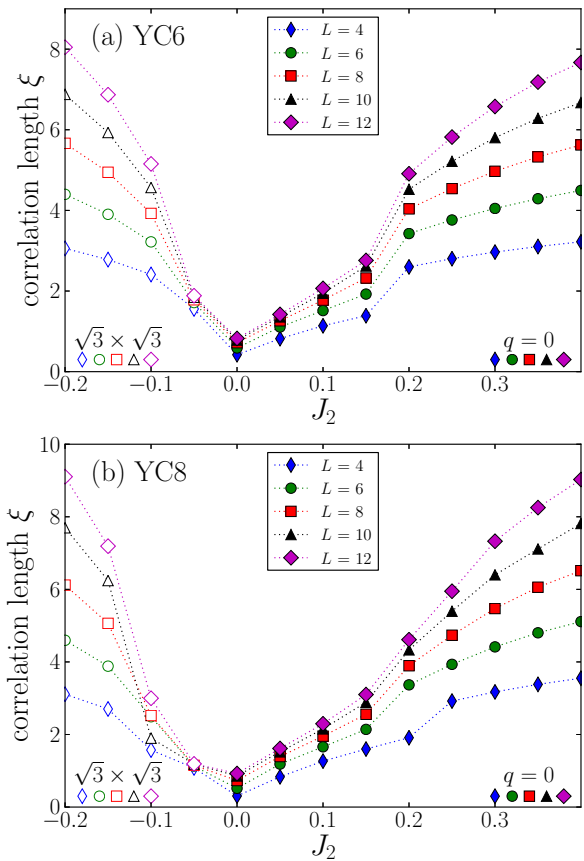}
\caption{
 (Color online) The magnetic correlation length $\xi$ calculated 
 from Eq.~\eqref{eq:corrlength} as a function of $J_{2}$ for (a) 
 the YC6 and (b) the YC8 cylinders at various cylinder lengths 
 $L$. For $J_{2}\leq-0.05$ the correlation length (empty symbols) 
 is calculated using $\mathbf{Q}=\mathbf{q}_K$ in~\eqref{eq:corrlength} 
 and measures the strength of the $\sqrt{3}\times\sqrt{3}$ magnetic 
 order. For $J_{2} \geq 0$ $\xi$ (full symbols) is defined 
 using $\mathbf{Q}=\mathbf{q}_M$ and it measures the strength of 
 the $q=0$ magnetic order.
}
\label{fig:corrlength}
\end{figure}

Figure~\ref{fig:corrlength} plots $\xi(\mathbf{Q},\mathbf{q}_{min})$ for 
the YC6 and YC8 cylinders ((a) and (b) in the Figure) and various cylinder 
lengths $L$. In the Figure we show $\xi({\mathbf{q}_K},\mathbf{q}_{min})$ (empty 
symbols) and $\xi({\mathbf{q}_M},\mathbf{q}_{min})$ (full symbols) in the region 
with $J_2<0$ and $J_2\ge 0$, respectively. The qualitative behavior is the same 
for both YC6 and YC8 cylinders. We obtain small correlation lengths with weak 
dependence on the cylinder length for $-0.1 \lesssim J_{2} \lesssim 0.15$. In 
particular, at $J_{2}=0$ both correlation lengths are of the order of the lattice 
constant, as expected in a spin liquid~\cite{Depenbrock2012}. This behavior reflects 
that of the order parameter $m^2_{\mathbf{Q}}$ (cf. Figure~\ref{fig:orderparameter}). 
Outside the disordered region the correlation lengths show an increasing trend as a 
function of the cylinder length $L$. For the extremal values $J_2=-0.2$ and $J_2=0.4$ 
considered in this work $\xi(\mathbf{Q},\mathbf{q}_{min})$ is of the order of the 
system size. 

\section{The spin triplet gaps}
\label{sec5}

Using $SU(2)$-invariant DMRG simulations we obtain the lowest-energy 
eigenstate in both the $S=0$ and $S=1$ sectors. We extrapolate their energies in the single-site DMRG truncation error to get the best ground state energy estimate. Subtracting the extrapolated energies we obtain the spin triplet gap $\Delta_t$. This is plotted in Figure~\ref{fig:gaps} for the YC6 and YC8 geometries and several cylinder lengths. Errorbars result from the extrapolation in the truncation error and are in many cases smaller than the symbol sizes. In both cases the gap 
shows the same qualitative behavior. There is a dome-shaped region for $-0.1
\lesssim J_{2}\lesssim 0.2$, with a weak dependence on $L$ and a peak at $J_{2}\simeq 0.1$. 
Remarkably, at the kagome point $J_2=0$ the triplet gap is almost independent of 
the system size, and its value $\Delta_t\approx 0.13$ is in perfect 
agreement with the result $\Delta_t=0.13(1)$ of Ref.~\onlinecite{Depenbrock2012}. 
A sharp dip is visible at $J_2\approx 0.2$ and for both geometries, which could suggest a 
phase transition between the spin-liquid and the $q=0$ ordered phase in the 
thermodynamic limit. A less pronounced feature is also visible at $J_2\approx -0.1$. 
For both $J_2\lesssim -0.1$ and $J_2\gtrsim 0.2$ $\Delta_t$ shows a strong dependence on 
the system size with a decreasing trend as a function of $L,W$, suggesting a vanishing 
behavior in the limit $L,W\to\infty$, as expected in a magnetically ordered phase.

\begin{figure}[h]
\includegraphics[width=.85\linewidth]{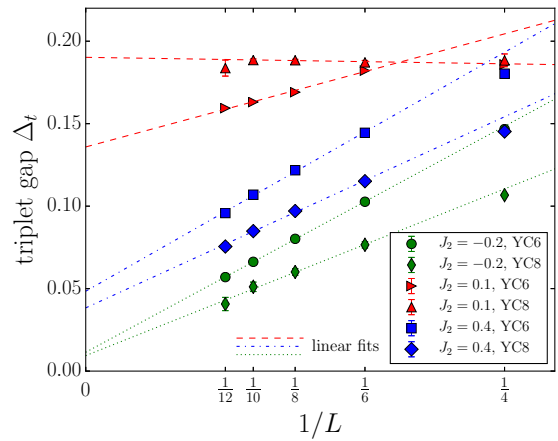}
\caption{
 (Color online) The spin triplet gap in the $J_{1}$-$J_{2}$ Heisenberg 
 model on the kagome lattice plotted versus $1/L$ for both YC6 and 
 YC8 cylinders. Data for $L=4,6,8,10,12$ and $J_2=-0.2,0.1,0.4$ are 
 shown in the figure. The lines denote the linear extrapolations to 
 the infinite cylinder limit. 
}
\label{fig:gaps-extrapolation-text}
\end{figure}

It is interesting to investigate the behavior of $\Delta_t$ in the limit $L\to\infty$, 
i.e., for infinitely long cylinders. This is illustrated in Figure~\ref{fig:gaps-extrapolation-text},  
plotting  $\Delta_t$ as a function of $1/L$ for $J_{2}=-0.2,0.1,0.4$ and both 
YC6 and YC8 cylinders. The dotted lines are the linear extrapolations to the infinite 
cylinder limit. The extrapolated gaps are shown in Figure~\ref{fig:gaps-extrapolated}. 
The triplet gap shows a peak at $J_{2}\simeq 0.1$ with a value of approximately 
$\Delta_t\approx 0.14$ for the YC6 and $\Delta_t\approx 0.18$ for the YC8 cylinder. 
It is interesting to observe that the maximum of the gap is not at $J_2\approx 0$, 
where the structure factor is featureless (cf. Figure~\ref{fig:orderparameter}). 
For larger $|J_2|$ the extrapolated gap exhibits decreasing behavior as a function of 
$J_2$. We should remark that, although the extrapolated gaps seem to vanish outside the 
disordered region, this should not be associated with the presence of Goldstone 
modes, as infinite long cylinders are expected to exhibit $1D$ behavior and no 
symmetry breaking. 

\section{Discussion}
\label{sec6}

Here we discuss the physical implications of the numerical results 
presented in section~\ref{sec3}, section~\ref{sec4}, and section~\ref{sec5}. 
We divide the discussion in three parts for different parameter ranges. 
First we consider the case $J_{2}\lesssim -0.1$, then $-0.1\lesssim J_{2}
\lesssim 0.2$, and finally $J_{2} \gtrsim 0.2$. 

\begin{figure}[t]
\includegraphics[width=.95\linewidth]{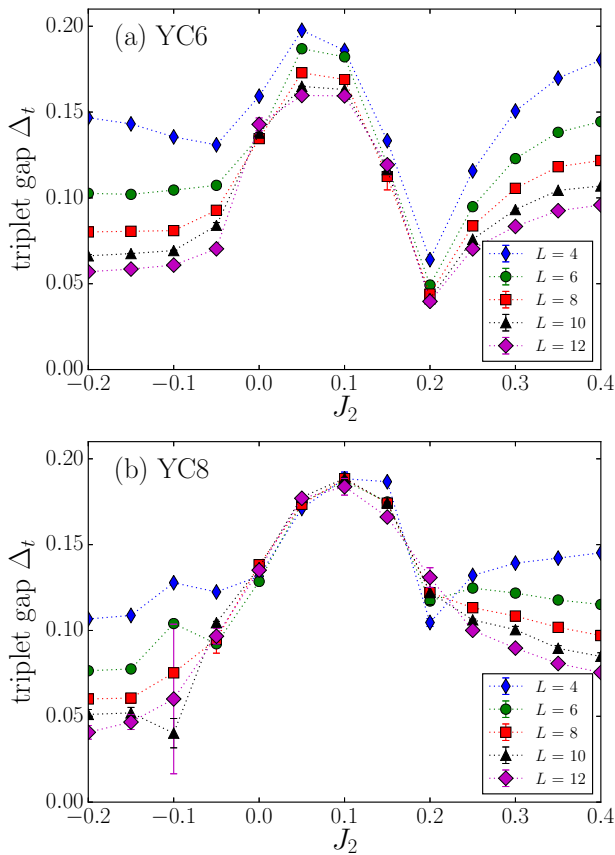}
\caption{
 (Color online) The spin triplet gap $\Delta_t$ of the $J_{1}$-$J_{2}$ 
 Heisenberg model on the kagome lattice as a function of next-nearest 
 neighbor interaction $J_{2}$. Spin gaps for the YC6 (see (a)) and 
 the YC8 cylinder (see (b)), and various cylinder lengths are shown. 
 The gaps are obtained by subtracting the energies of the lowest-energy 
 states in the $S=0$ and $S=1$ symmetry sectors, which can be directly 
 accessed by $SU(2)$-symmetric DMRG simulations.
}
\label{fig:gaps}
\end{figure}

\begin{figure}[t]
\includegraphics[width=.95\linewidth]{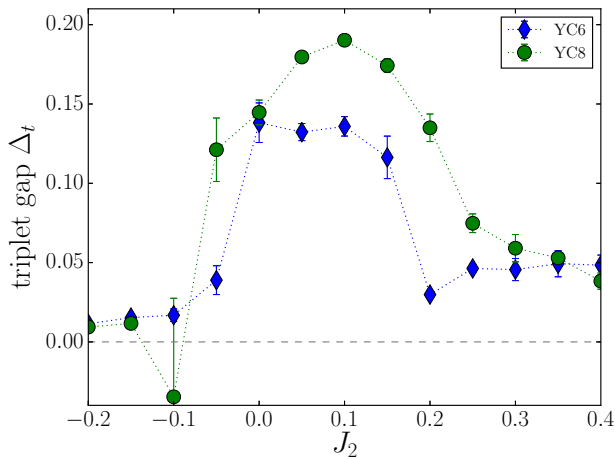}
\caption{
 The spin triplet gap $\Delta_t$ in the $J_{1}$-$J_{2}$ 
 Heisenberg model: Extrapolations to the infinitely long cylinder 
 limit. We show $\Delta_t$ for both the YC6 and the YC8 cylinders as a 
 function of the next-nearest neighbor coupling $J_{2}$. 
 Error bars result from the extrapolation in the cylinder length  
 (see Figure~\ref{fig:gaps-extrapolation-text}). 
}
\label{fig:gaps-extrapolated}
\end{figure}

\paragraph{$J_{2}\lesssim -0.1$.}
The static spin structure factor at $J_{2}\approx-0.2$ (see Figure~\ref{fig:structurefactor} 
(i)) exhibits sharp peaks at the $K$-points of the extended Brillouin zone. The peak 
positions are in agreement with what is expected for the classical 
$\sqrt{3}\times\sqrt{3}$ order. Moreover, the DMRG data suggest a sudden increase of  
the antiferromagnetic order parameter $m^2_{\mathbf{Q}}$ with increasing $|J_{2}|$. 
The corresponding spin-spin correlation length is of the order of the 
system size, and it increases upon increasing $|J_2|$. This could suggest   
magnetic order of the $\sqrt{3}\times\sqrt{3}$ type in the thermodynamic limit. This 
is also weakly confirmed by the behavior of the triplet gap $\Delta_t$. We numerically 
observe that $\Delta_t$ decreases upon increasing  $L$ and $W$ for $J_2\lesssim -0.1$, 
which is consistent with a vanishing behavior in the $2D$ limit (cf. Figure~\ref{fig:gaps} 
and Figure~\ref{fig:gaps-extrapolated}), as expected in a magnetically ordered phase, 
due to the presence of the Goldstone modes. 

\paragraph{$-0.1\lesssim J_{2} \lesssim 0.2$.}
In this region we observe a dome-shaped triplet gap. For both the YC6 and YC8 
geometries  the DMRG data support a finite gap in the infinite cylinder limit (see 
Figure~\ref{fig:gaps-extrapolated}), excluding the presence of magnetic order. 
Interestingly, for the YC8 cylinders this gap is almost independent of the cylinder 
length. The structure factor is almost featureless (cf.~Figure~\ref{fig:structurefactor}),  
although some peaks at the $M$-points of the extended Brillouin zone are visible, 
signalling the onset of the $q=0$ order at larger $J_2$. The spin-spin correlation lengths 
for both the $\sqrt{3}\times \sqrt{3}$ and $q=0$ magnetic order are of the order of the 
lattice constant. These results confirm earlier DMRG studies performed at $J_{2}=0$~\cite{Yan2011,
Depenbrock2012}, in agreement with an extended $Z_2$ spin liquid region around $J_2=0$.
Notice that our data does not support an algebraic $U(1)$ spin liquid, which would 
imply a vanishing spin gap, in contrast with what has been found recently by variational 
Monte Carlo methods~\cite{Iqbal2012}. Also, from the present data we cannot exclude a 
transition from the $Z_{2}$ spin liquid to a valence bond crystal ($VBC$) for small 
ferromagnetic $J_2$, as it was reported in Ref.~\onlinecite{Iqbal2012}. Notice that the
breaking of the lattice symmetry is hard to detect~\cite{Sandvik2012} using the cylinder geometry. 
In order to detect the $VBC$ phase it would be useful to study the dimer-dimer correlation 
function $\langle D_{i}^\alpha D_{j}^\beta\rangle$, where $D_i^\alpha\equiv S_{\mathbf{r}_i}
\cdot S_{\mathbf{r}_i+\alpha}$, with $\alpha=\hat x,\hat y$, and the 
corresponding structure factor $S_d^{\alpha,\beta}(\mathbf{q})\equiv 1/N\sum_{i,j}
e^{i\mathbf{q}(\mathbf{r}_i-\mathbf{r}_j)}\langle D_{i}^\alpha D_j^\beta\rangle$.
Moreover, it would be interesting to calculate the topological entanglement 
entropy $\gamma$, which is expected to be zero in the $VBC$ phase, while it is 
$\gamma=\log(2)$ in the $Z_{2}$ spin liquid phase. However, this would require larger 
cylinders in order to perform a precise finite-size scaling analysis of the von 
Neumann entropy.

\paragraph{$ 0.2 \lesssim J_{2}$}
We find sharp peaks in the static spin structure factor (cf. Figure~\ref{fig:structurefactor} 
(iv)) at the $M$-points of the extended Brillouin zone. This is in agreement with what 
is expected for the $q=0$ magnetic order. The triplet gap exhibits a decreasing behavior 
upon increasing $W$ and $L$. Correspondingly, the spin-spin correlation 
length rapidly increases with $J_2$ (cf. Figure~\ref{fig:corrlength}).

\section{Entanglement spectroscopy in the $q=0$ phase}
\label{sec7}

Given a spatial bipartition of the cylinder in parts $A$ and $B$, the 
so-called entanglement spectrum (ES) levels~\cite{Li2008} $\{\xi_{i}\}$ are 
constructed from the Schmidt decomposition of the ground-state wavefunction 
$|\psi\rangle$ as 
\begin{align}
|\psi\rangle = \sum_{i}e^{\xi_{i}/2} |\psi_{i}^{A}\rangle \otimes |
\psi_{i}^{B}\rangle,
\end{align}
where $|\psi^{A(B)}\rangle$ form an orthonormal basis set for subsystem 
$A(B)$. Alternatively, the ES can be thought of as  the spectrum of an 
effective entanglement Hamiltonian $\mathcal{H}_{E}$ that is defined as
\begin{align}
\mathcal{H}_{E}\equiv\exp (-\rho_{A}),
\end{align}
where $\rho_{A}$ is the reduced density matrix of subsystem $A$.
Since the DMRG algorithm works directly in the Schmidt basis the ES is 
available essentially for free during a ground state simulation and provides 
another useful tool to characterize the properties of the ground state. 

\begin{figure}[t]
\includegraphics[width=.85\linewidth]{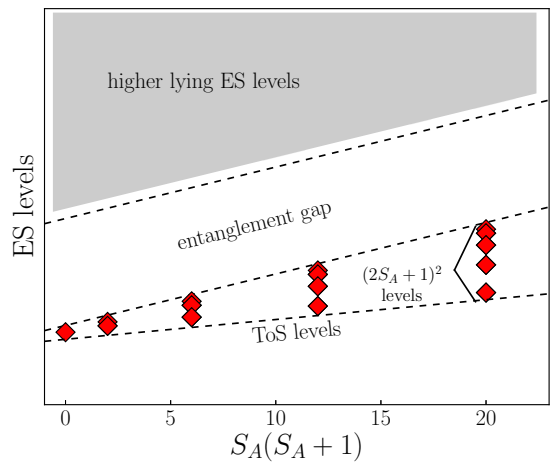}
\caption{(Color online) Cartoon of the expected structure of the 
 entanglement spectrum (ES) in the $q=0$ magnetically ordered phase 
 in the $J_1$-$J_2$ Heisenberg model on the kagome lattice. ES 
 levels are plotted versus $S_A(S_A+1)$, with $S_A$ being the total 
 spin in subsystem $A$. The rhombi denote the ES levels displaying 
 the tower-of-states structure. The number of tower-of-states levels 
 in each sector with fixed $S_A$ is given as $(2S_A+1)^2$. These 
 are divided from the higher-lying levels by an entanglement 
 ``gap''. 
}
\label{fig:tos_schematic}
\end{figure}

It has been proposed recently~\cite{Metlitski2011} that in a model that 
breaks a continuous symmetry in the thermodynamic limit the low-lying part of 
the ground-state entanglement spectrum (ES) exhibits the tower-of-states 
structure, which describes the finite-size energy spectrum of the model. In 
particular, for a spin model that fully breaks the $SU(2)$ symmetry, 
many features of the low-lying ES levels can be understood in terms of the    
entanglement Hamiltonian  
\begin{align}
\label{ToS_ham}
\mathcal{H}_{E}\propto\frac{\mathbf{S}_{A}^{2}}{W}+\cdots, 
\end{align}
where $\mathbf{S}_A$ is the total spin in subsystem $A$ and $W\sim\sqrt{N}$ 
the cylinder width (cf. Figure~\ref{fig:lattice}). The low-lying spectrum 
of~\eqref{ToS_ham} is shown schematically in Figure~\ref{fig:tos_schematic}, 
plotting ES levels versus $S_A(S_A+1)$. In each sector with fixed $S_A$ 
there are $(2S_A+1)^2$ levels (rhombi in the Figure) forming the 
tower-of-states, which are divided from higher-lying levels by an 
entanglement gap. The tower-of-states levels exhibit linear behavior with 
respect to $S_A(S_A+1)$. Notice that, although~\eqref{ToS_ham} gives $(2S_A+1)^2$ 
degenerate levels in each spin sector, this degeneracy is in general lifted, 
as shown in Figure~\ref{fig:tos_schematic}. The correspondence between ES and tower-of-states 
has been numerically verified in the $J_1$-$J_2$ KHA in Ref.~\onlinecite{Kolley2013} 
for $J_{2}=-1.0$, i.e. deep in the $\sqrt{3} \times \sqrt{3}$ ordered phase. 

Notice that both the $\sqrt{3}\times \sqrt{3}$ and the $q=0$ ordering patterns 
correspond to full breaking of the $SU(2)$ symmetry (see Figure~\ref{fig:classical-order}), 
as they contain three ferromagnetic sublattices. As a consequence, deep  
in the $q=0$ phase one should expect the same tower-of-states structure shown in 
Figure~\ref{fig:tos_schematic} in the ES. However, here we provide numerical 
evidence that the identification of the correct tower-of-states depends on the 
choice of the bipartition, at least for small system sizes. 

\begin{center}
\begin{figure*}[htb]
\includegraphics[width=1.\linewidth]{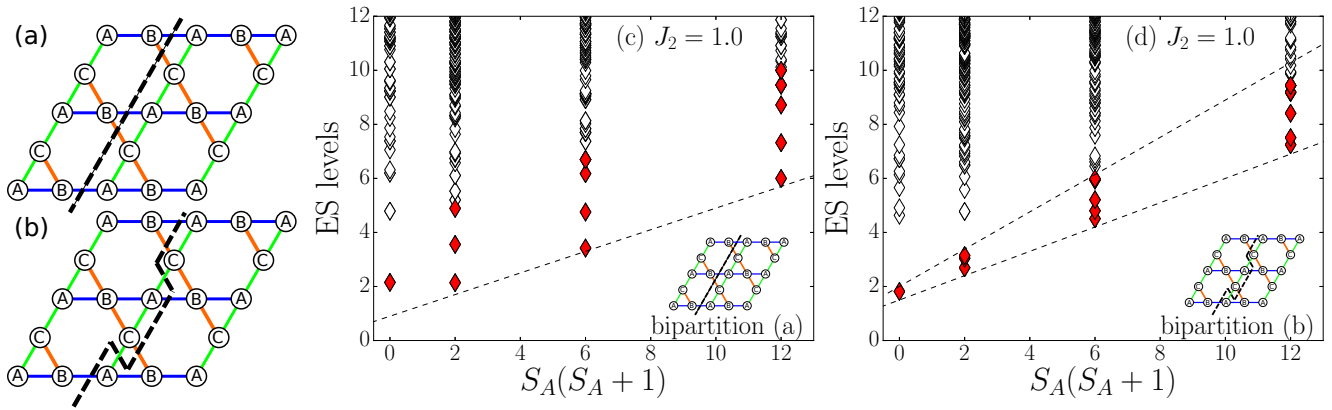}
\caption{
 (Color online) Spin resolved entanglement spectra obtained from the 
 ground state wavefunction of the $J_{1}$-$J_{2}$ Heisenberg model on 
 the kagome lattice at $J_{2}=1.0$. Data are for a cylinder with $4
 \times 12$ unit cells. Left: schematic representation of the $q=0$ 
 state on the kagome lattice. $A$, $B$, and $C$ denote the three 
 ferromagnetic sublattices. Bonds connecting spins in different 
 sublattices are are shown with different colors. The black dashed 
 line marks the cut defining the bipartition used to calculate the 
 entanglement spectrum. Two possible cuts are shown: (a) The cut 
 crosses only $A$-$B$ and $B$-$C$ bonds, (b) the cut crosses 
 $A$-$B$, $B$-$C$, and $A$-$C$ bonds. Right: The entanglement  
 spectra obtained from the bipartitions shown in (a) and (b), 
 plotted versus $S_A(S_A+1)$, with $S_A$ being the total spin in 
 subsystem $A$. Each symbol corresponds to a $2S_{A}+1$ degenerate 
 multiplet of levels. The red symbols denote the lowest $(
 2S_{A}+1)^{2}$ levels. Notice that deviations from the expected 
 tower-of-states structure (cf. Figure~\ref{fig:tos_schematic}) 
 are large using the bipartition shown in (a). 
}
\label{fig:es_J1}
\end{figure*}
\end{center}

This is illustrated in Figure~\ref{fig:es_J1} plotting the half-system ES for a 
kagome cylinder with $4\times12$ unit cells (YC8 geometry) at $J_{2}=1.0$ and 
for two different bipartitions. The bipartitions are shown in (a) and (b): The 
three ferromagnetic sublattices forming the $q=0$ state (cf. Figure~\ref{fig:lattice}) 
are denoted as $A$, $B$, $C$, bonds connecting spins on different sublattices are 
shown with different colors. While (a) corresponds to a straight cut, (b) has a zigzag 
structure. One should observe that the straight cut crosses only $B$-$C$ and $A$-$B$ 
bonds, whereas all the three types of bonds ($A$-$B$, $B$-$C$, and $A$-$C$) are 
crossed by the zigzag cut in (b). This suggests that the straight cut might not 
capture the quantum correlations between sublattices $A$ and $C$. Notice that for 
the $\sqrt{3}\times\sqrt{3}$ state this is not the case as the straight cut 
would cross all the three different types of bonds. The difference between the two 
cuts is reflected in the corresponding entanglement spectra. 

The ES obtained using the straight cut (a) is reported in Figure~\ref{fig:es_J1} 
(c). The ES levels are plotted versus $S_A(S_A+1)$. Full symbols denote the lowest 
$(2S_A+1)^2$ levels in each spin sector. Strong deviations from the expected picture 
in Figure~\ref{fig:tos_schematic} are visible. In particular, no gap between the 
tower-of-states levels and the rest of the spectrum is visible. 
Better agreement with Figure~\ref{fig:tos_schematic} is found using the zigzag cut, 
as it is clear from Figure~\ref{fig:es_J1} (d). For instance, the low-lying levels 
now show a clear linear behavior with respect to $S_{A}(S_{A}+1)$. Moreover, in the 
$S_{A}=0$ and $S_{A}=1$ sectors the tower-of-states levels are well separated from 
higher-lying levels by an entanglement gap, although this  
becomes smaller for $S_{A}=2$, when the low-lying levels start mixing with the 
rest of the spectrum. Finally, we should mention that, despite the numerical evidence 
in Figure~\ref{fig:es_J1}, within the available system sizes we cannot exclude that 
the difference between the ES in (c) (d) disappears considering larger cylinders. 

\section{Conclusion}\label{sec:conclusion}
 
We performed an extensive DMRG study of the ground-state phase diagram of the 
$J_{1}$-$J_{2}$ Heisenberg model on kagome cylinders. We restricted ourselves 
to $J_1=1$, considering both antiferromagnetic and ferromagnetic $J_2$. In 
particular, we investigated the behavior of the model around the pure kagome 
point at $J_2=0$. To this purpose, we monitored the behavior of the spin triplet 
gap, the static structure factor, and the magnetic correlation length, as 
a function of $J_2$. We should remark that our results are based on 
finite-size cylinders. Strong finite-size effects do not allow us to provide 
conclusive results about the phase diagram of the model in the thermodynamic 
limit. 

By comparing the finite-size behaviors of the spin gap, the structure factor, and 
the correlation lengths, we found numerical evidence suggesting that the ground 
state of the model displays magnetic order for $J_2\lesssim -0.1$ and $J_2\gtrsim 0.2$. 
Precisely, for $J_2\lesssim -0.1$ the structure factor exhibits sharp peaks at the 
$K$-points of the extended Brillouin zone, in agreement with what is expected 
for the classical $\sqrt{3}\times\sqrt{3}$ state, whereas at $J_2\gtrsim0.2$ one 
observes peaks at the $M$-points, which signal the $q=0$ magnetic pattern. 
In both cases the correlation lengths associated with the two structures show a 
rapid increase upon increasing $|J_2|$ and the system size. Correspondingly, 
the triplet gap decreases, suggesting a vanishing gap in the thermodynamic limit. 
Within the system sizes accessible to the simulations our results 
are consistent with the presence of a magnetically disordered phase for 
$-0.1 \lesssim J_{2} \lesssim 0.2$, which is compatible with spin-liquid 
behavior~\cite{Jiang2012}. In this region the spin gap shows a weaker  
dependence on the cylinder size. Moreover, the DMRG data support a finite 
gap for infinitely long cylinders. The static structure factor is featureless at 
the $J_2=0$ point, and it exhibits not very pronounced structures in the whole 
region $-0.1 \lesssim J_{2}\lesssim 0.2$. The magnetic correlation lengths 
associated with the $\sqrt{3}\times\sqrt{3}$ and the $q=0$ order are of the order 
of the lattice unit.

As a final point, we investigated the structure of the ground state entanglement 
spectrum (ES) in the $q=0$ ordered phase. We found that the identification of 
the tower-of-states structure, which is associated with the $SU(2)$ symmetry 
breaking in the thermodynamic limit, depends dramatically on the choice of the 
spatial bipartition of the state, at least for small system sizes. 

Recently, we became aware of two related works. In Ref.~\onlinecite{Gong2014b} a DMRG study of the phase diagram of the $J_{1}$-$J_{2}$-$J_{3}$-Heisenberg model on the kagome lattice is performed and in Ref.~\onlinecite{Iqbal2015} the phase diagram of the $J_{1}$-$J_{2}$-Heisenberg model on the kagome lattice is studied with a variational Monte Carlo method. The results of both works are in qualitative agreement with the ones reported in this paper. 

\begin{acknowledgments}
U. S. thanks Ronny Thomale for useful discussions. U. S. and V. A. acknowledge 
funding by DFG through NIM and SFB/TR 12.
\end{acknowledgments}

\appendix*
\section{Independence on boundary conditions}

%
\begin{figure}[b]
\includegraphics[width=1.\linewidth]{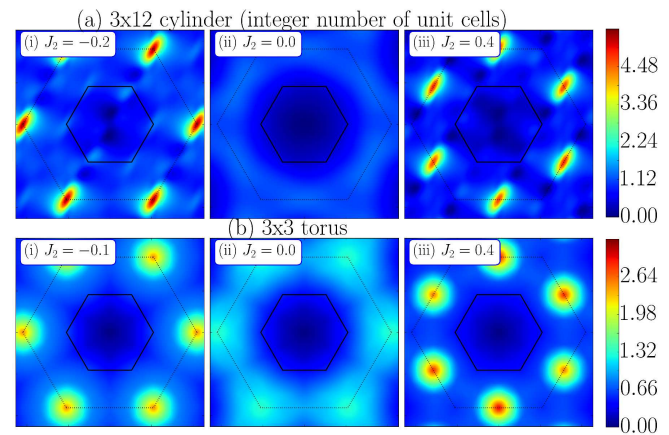}
\caption{(Color online) The static spin structure factor 
 $S(\mathbf{q})$ obtained from ground-state DMRG simulations 
 of the $J_{1}$-$J_{2}$ Heisenberg model (a) on kagome cylinders with integer number of $3\times 12$ unit cells and (b) on small tori  with $3\times 3$ unit cells for different values of $J_{2}$. The solid and the dotted lines show the first and the extended 
 Brillouin zones, respectively. The spin correlations are qualitatively the same as in Fig.~\ref{fig:structurefactor} in the main body of the text for all $J_{2}$ values and boundary conditions.
}
\label{fig:boundaries}
\end{figure}

In this work we used a lattice geometry with a non-integer number of unit cells 
(cf. Fig.~\ref{fig:lattice}). 
Here we show that this does not affect the phase diagram presented in Fig.~\ref{fig:structurefactor}. 
We also investigate the effect of boundary conditions discussing the structure factor for the 
$J_1$-$J_2$ Heisenberg model on kagome tori. 
In Fig.~\ref{fig:boundaries} we present the static spin structure factors for lattices with 
an integer number of unit cells at $J_{2}=-0.2, 0.0, 0.4$ as well as for fully periodic 
small tori at $J_{2}=-0.1, 0.0, 0.4$. For all the lattice geometries we find antiferromagnetic 
correlations corresponding to the $\sqrt{3}\times\sqrt{3}$ state at $J_{2}=-0.2$ and $J_{2}=-0.1$, 
and antiferromagnetic correlations corresponding to the $q=0$ state at $J_{2}=0.4$. At $J_{2}=0.0$ 
the structure factor for the cylinder geometry with integer number of unit cell is structureless 
while for the small torus the structure factor shows slightly enhanced correlations at the $K$-points 
of the extended Brillouin zone.

\bibliographystyle{apsrev}
\bibliography{/home/f/F.Kolley/Documents/myLib} 

\begin{thebibliography}{59}
\expandafter\ifx\csname natexlab\endcsname\relax\def\natexlab#1{#1}\fi
\expandafter\ifx\csname bibnamefont\endcsname\relax
  \def\bibnamefont#1{#1}\fi
\expandafter\ifx\csname bibfnamefont\endcsname\relax
  \def\bibfnamefont#1{#1}\fi
\expandafter\ifx\csname citenamefont\endcsname\relax
  \def\citenamefont#1{#1}\fi
\expandafter\ifx\csname url\endcsname\relax
  \def\url#1{\texttt{#1}}\fi
\expandafter\ifx\csname urlprefix\endcsname\relax\def\urlprefix{URL }\fi
\providecommand{\bibinfo}[2]{#2}
\providecommand{\eprint}[2][]{\url{#2}}

\bibitem[{\citenamefont{Zeng and Elser}(1990)}]{Zeng1990}
\bibinfo{author}{\bibfnamefont{C.}~\bibnamefont{Zeng}} \bibnamefont{and}
  \bibinfo{author}{\bibfnamefont{V.}~\bibnamefont{Elser}},
  \bibinfo{journal}{Phys. Rev. B} \textbf{\bibinfo{volume}{42}},
  \bibinfo{pages}{8436} (\bibinfo{year}{1990}).

\bibitem[{\citenamefont{Marston and Zeng}(1991)}]{Marston1991}
\bibinfo{author}{\bibfnamefont{J.~B.} \bibnamefont{Marston}} \bibnamefont{and}
  \bibinfo{author}{\bibfnamefont{C.}~\bibnamefont{Zeng}}, \bibinfo{journal}{J.
  Appl. Phys.} \textbf{\bibinfo{volume}{69}}, \bibinfo{pages}{5962}
  (\bibinfo{year}{1991}).

\bibitem[{\citenamefont{Hastings}(2000)}]{Hastings2000}
\bibinfo{author}{\bibfnamefont{M.~B.} \bibnamefont{Hastings}},
  \bibinfo{journal}{Phys. Rev. B} \textbf{\bibinfo{volume}{63}},
  \bibinfo{pages}{014413} (\bibinfo{year}{2000}),
  \urlprefix\url{http://link.aps.org/doi/10.1103/PhysRevB.63.014413}.

\bibitem[{\citenamefont{Nikolic and Senthil}(2003)}]{Nikolic2003}
\bibinfo{author}{\bibfnamefont{P.}~\bibnamefont{Nikolic}} \bibnamefont{and}
  \bibinfo{author}{\bibfnamefont{T.}~\bibnamefont{Senthil}},
  \bibinfo{journal}{Phys. Rev. B} \textbf{\bibinfo{volume}{68}},
  \bibinfo{pages}{214415} (\bibinfo{year}{2003}),
  \urlprefix\url{http://link.aps.org/doi/10.1103/PhysRevB.68.214415}.

\bibitem[{\citenamefont{Singh and Huse}(2007)}]{Singh2007}
\bibinfo{author}{\bibfnamefont{R.~R.~P.} \bibnamefont{Singh}} \bibnamefont{and}
  \bibinfo{author}{\bibfnamefont{D.~A.} \bibnamefont{Huse}},
  \bibinfo{journal}{Phys. Rev. B} \textbf{\bibinfo{volume}{76}},
  \bibinfo{pages}{180407} (\bibinfo{year}{2007}),
  \urlprefix\url{http://link.aps.org/doi/10.1103/PhysRevB.76.180407}.

\bibitem[{\citenamefont{Singh and Huse}(2008)}]{Singh2008}
\bibinfo{author}{\bibfnamefont{R.~R.~P.} \bibnamefont{Singh}} \bibnamefont{and}
  \bibinfo{author}{\bibfnamefont{D.~A.} \bibnamefont{Huse}},
  \bibinfo{journal}{Phys. Rev. B} \textbf{\bibinfo{volume}{77}},
  \bibinfo{pages}{144415} (\bibinfo{year}{2008}),
  \urlprefix\url{http://link.aps.org/doi/10.1103/PhysRevB.77.144415}.

\bibitem[{\citenamefont{Iqbal et~al.}(2012)\citenamefont{Iqbal, Becca, and
  Poilblanc}}]{Iqbal2012}
\bibinfo{author}{\bibfnamefont{Y.}~\bibnamefont{Iqbal}},
  \bibinfo{author}{\bibfnamefont{F.}~\bibnamefont{Becca}}, \bibnamefont{and}
  \bibinfo{author}{\bibfnamefont{D.}~\bibnamefont{Poilblanc}},
  \bibinfo{journal}{New J. Phys.} \textbf{\bibinfo{volume}{14}},
  \bibinfo{pages}{115031} (\bibinfo{year}{2012}).

\bibitem[{\citenamefont{Kalmeyer and Laughlin}(1989)}]{Kalmeyer1989}
\bibinfo{author}{\bibfnamefont{V.}~\bibnamefont{Kalmeyer}} \bibnamefont{and}
  \bibinfo{author}{\bibfnamefont{R.~B.} \bibnamefont{Laughlin}},
  \bibinfo{journal}{Phys. Rev. B} \textbf{\bibinfo{volume}{39}},
  \bibinfo{pages}{11879} (\bibinfo{year}{1989}),
  \urlprefix\url{http://link.aps.org/doi/10.1103/PhysRevB.39.11879}.

\bibitem[{\citenamefont{Sachdev}(1992)}]{Sachdev1992}
\bibinfo{author}{\bibfnamefont{S.}~\bibnamefont{Sachdev}},
  \bibinfo{journal}{Phys. Rev. B} \textbf{\bibinfo{volume}{45}},
  \bibinfo{pages}{12377} (\bibinfo{year}{1992}),
  \urlprefix\url{http://link.aps.org/doi/10.1103/PhysRevB.45.12377}.

\bibitem[{\citenamefont{Yang et~al.}(1993)\citenamefont{Yang, Warman, and
  Girvin}}]{Yang1993}
\bibinfo{author}{\bibfnamefont{K.}~\bibnamefont{Yang}},
  \bibinfo{author}{\bibfnamefont{L.~K.} \bibnamefont{Warman}},
  \bibnamefont{and} \bibinfo{author}{\bibfnamefont{S.~M.}
  \bibnamefont{Girvin}}, \bibinfo{journal}{Phys. Rev. Lett.}
  \textbf{\bibinfo{volume}{70}}, \bibinfo{pages}{2641} (\bibinfo{year}{1993}),
  \urlprefix\url{http://link.aps.org/doi/10.1103/PhysRevLett.70.2641}.

\bibitem[{\citenamefont{Misguich et~al.}(2002)\citenamefont{Misguich, Serban,
  and Pasquier}}]{Misguich2002}
\bibinfo{author}{\bibfnamefont{G.}~\bibnamefont{Misguich}},
  \bibinfo{author}{\bibfnamefont{D.}~\bibnamefont{Serban}}, \bibnamefont{and}
  \bibinfo{author}{\bibfnamefont{V.}~\bibnamefont{Pasquier}},
  \bibinfo{journal}{Phys. Rev. Lett.} \textbf{\bibinfo{volume}{89}},
  \bibinfo{pages}{137202} (\bibinfo{year}{2002}),
  \urlprefix\url{http://link.aps.org/doi/10.1103/PhysRevLett.89.137202}.

\bibitem[{\citenamefont{Wang and Vishwanath}(2006)}]{Wang2006}
\bibinfo{author}{\bibfnamefont{F.}~\bibnamefont{Wang}} \bibnamefont{and}
  \bibinfo{author}{\bibfnamefont{A.}~\bibnamefont{Vishwanath}},
  \bibinfo{journal}{Phys. Rev. B} \textbf{\bibinfo{volume}{74}},
  \bibinfo{pages}{174423} (\bibinfo{year}{2006}),
  \urlprefix\url{http://link.aps.org/doi/10.1103/PhysRevB.74.174423}.

\bibitem[{\citenamefont{Ryu et~al.}(2007)\citenamefont{Ryu, Motrunich, Alicea,
  and Fisher}}]{Ryu2007}
\bibinfo{author}{\bibfnamefont{S.}~\bibnamefont{Ryu}},
  \bibinfo{author}{\bibfnamefont{O.~I.} \bibnamefont{Motrunich}},
  \bibinfo{author}{\bibfnamefont{J.}~\bibnamefont{Alicea}}, \bibnamefont{and}
  \bibinfo{author}{\bibfnamefont{M.~P.~A.} \bibnamefont{Fisher}},
  \bibinfo{journal}{Phys. Rev. B} \textbf{\bibinfo{volume}{75}},
  \bibinfo{pages}{184406} (\bibinfo{year}{2007}),
  \urlprefix\url{http://link.aps.org/doi/10.1103/PhysRevB.75.184406}.

\bibitem[{\citenamefont{Ran et~al.}(2007)\citenamefont{Ran, Hermele, Lee, and
  Wen}}]{Ran2007}
\bibinfo{author}{\bibfnamefont{Y.}~\bibnamefont{Ran}},
  \bibinfo{author}{\bibfnamefont{M.}~\bibnamefont{Hermele}},
  \bibinfo{author}{\bibfnamefont{P.~A.} \bibnamefont{Lee}}, \bibnamefont{and}
  \bibinfo{author}{\bibfnamefont{X.-G.} \bibnamefont{Wen}},
  \bibinfo{journal}{Phys. Rev. Lett.} \textbf{\bibinfo{volume}{98}},
  \bibinfo{pages}{117205} (\bibinfo{year}{2007}),
  \urlprefix\url{http://link.aps.org/doi/10.1103/PhysRevLett.98.117205}.

\bibitem[{\citenamefont{Hermele et~al.}(2008)\citenamefont{Hermele, Ran, Lee,
  and Wen}}]{Hermele2008}
\bibinfo{author}{\bibfnamefont{M.}~\bibnamefont{Hermele}},
  \bibinfo{author}{\bibfnamefont{Y.}~\bibnamefont{Ran}},
  \bibinfo{author}{\bibfnamefont{P.~A.} \bibnamefont{Lee}}, \bibnamefont{and}
  \bibinfo{author}{\bibfnamefont{X.-G.} \bibnamefont{Wen}},
  \bibinfo{journal}{Phys. Rev. B} \textbf{\bibinfo{volume}{77}},
  \bibinfo{pages}{224413} (\bibinfo{year}{2008}),
  \urlprefix\url{http://link.aps.org/doi/10.1103/PhysRevB.77.224413}.

\bibitem[{\citenamefont{Jiang et~al.}(2008)\citenamefont{Jiang, Weng, and
  Sheng}}]{Jiang2008}
\bibinfo{author}{\bibfnamefont{H.~C.} \bibnamefont{Jiang}},
  \bibinfo{author}{\bibfnamefont{Z.~Y.} \bibnamefont{Weng}}, \bibnamefont{and}
  \bibinfo{author}{\bibfnamefont{D.~N.} \bibnamefont{Sheng}},
  \bibinfo{journal}{Phys. Rev. Lett.} \textbf{\bibinfo{volume}{101}},
  \bibinfo{pages}{117203} (\bibinfo{year}{2008}),
  \urlprefix\url{http://link.aps.org/doi/10.1103/PhysRevLett.101.117203}.

\bibitem[{\citenamefont{Iqbal et~al.}(2011)\citenamefont{Iqbal, Becca, and
  Poilblanc}}]{Iqbal2011}
\bibinfo{author}{\bibfnamefont{Y.}~\bibnamefont{Iqbal}},
  \bibinfo{author}{\bibfnamefont{F.}~\bibnamefont{Becca}}, \bibnamefont{and}
  \bibinfo{author}{\bibfnamefont{D.}~\bibnamefont{Poilblanc}},
  \bibinfo{journal}{Phys. Rev. B} \textbf{\bibinfo{volume}{84}},
  \bibinfo{pages}{020407} (\bibinfo{year}{2011}),
  \urlprefix\url{http://link.aps.org/doi/10.1103/PhysRevB.84.020407}.

\bibitem[{\citenamefont{Lu et~al.}(2011)\citenamefont{Lu, Ran, and
  Lee}}]{Lu2011}
\bibinfo{author}{\bibfnamefont{Y.-M.} \bibnamefont{Lu}},
  \bibinfo{author}{\bibfnamefont{Y.}~\bibnamefont{Ran}}, \bibnamefont{and}
  \bibinfo{author}{\bibfnamefont{P.~A.} \bibnamefont{Lee}},
  \bibinfo{journal}{Phys. Rev. B} \textbf{\bibinfo{volume}{83}},
  \bibinfo{pages}{224413} (\bibinfo{year}{2011}),
  \urlprefix\url{http://link.aps.org/doi/10.1103/PhysRevB.83.224413}.

\bibitem[{\citenamefont{Huh et~al.}(2011)\citenamefont{Huh, Punk, and
  Sachdev}}]{Huh2011}
\bibinfo{author}{\bibfnamefont{Y.}~\bibnamefont{Huh}},
  \bibinfo{author}{\bibfnamefont{M.}~\bibnamefont{Punk}}, \bibnamefont{and}
  \bibinfo{author}{\bibfnamefont{S.}~\bibnamefont{Sachdev}},
  \bibinfo{journal}{Phys. Rev. B} \textbf{\bibinfo{volume}{84}},
  \bibinfo{pages}{094419} (\bibinfo{year}{2011}),
  \urlprefix\url{http://link.aps.org/doi/10.1103/PhysRevB.84.094419}.

\bibitem[{\citenamefont{Messio et~al.}(2012)\citenamefont{Messio, Bernu, and
  Lhuillier}}]{Messio2012}
\bibinfo{author}{\bibfnamefont{L.}~\bibnamefont{Messio}},
  \bibinfo{author}{\bibfnamefont{B.}~\bibnamefont{Bernu}}, \bibnamefont{and}
  \bibinfo{author}{\bibfnamefont{C.}~\bibnamefont{Lhuillier}},
  \bibinfo{journal}{Phys. Rev. Lett.} \textbf{\bibinfo{volume}{108}},
  \bibinfo{pages}{207204} (\bibinfo{year}{2012}),
  \urlprefix\url{http://link.aps.org/doi/10.1103/PhysRevLett.108.207204}.

\bibitem[{\citenamefont{He et~al.}(2014{\natexlab{a}})\citenamefont{He, Sheng,
  and Chen}}]{He2014b}
\bibinfo{author}{\bibfnamefont{Y.-C.} \bibnamefont{He}},
  \bibinfo{author}{\bibfnamefont{D.~N.} \bibnamefont{Sheng}}, \bibnamefont{and}
  \bibinfo{author}{\bibfnamefont{Y.}~\bibnamefont{Chen}},
  \bibinfo{journal}{Phys. Rev. Lett.} \textbf{\bibinfo{volume}{112}},
  \bibinfo{pages}{137202} (\bibinfo{year}{2014}{\natexlab{a}}),
  \urlprefix\url{http://link.aps.org/doi/10.1103/PhysRevLett.112.137202}.

\bibitem[{\citenamefont{He et~al.}(2014{\natexlab{b}})\citenamefont{He, Sheng,
  and Chen}}]{He2014c}
\bibinfo{author}{\bibfnamefont{Y.-C.} \bibnamefont{He}},
  \bibinfo{author}{\bibfnamefont{D.~N.} \bibnamefont{Sheng}}, \bibnamefont{and}
  \bibinfo{author}{\bibfnamefont{Y.}~\bibnamefont{Chen}},
  \bibinfo{journal}{Phys. Rev. B} \textbf{\bibinfo{volume}{89}},
  \bibinfo{pages}{075110} (\bibinfo{year}{2014}{\natexlab{b}}),
  \urlprefix\url{http://link.aps.org/doi/10.1103/PhysRevB.89.075110}.

\bibitem[{\citenamefont{{Gong Shou-Shu} et~al.}(2014)\citenamefont{{Gong
  Shou-Shu}, {Zhu Wei}, and {Sheng D. N.}}}]{Gong2014a}
\bibinfo{author}{\bibnamefont{{Gong Shou-Shu}}},
  \bibinfo{author}{\bibnamefont{{Zhu Wei}}}, \bibnamefont{and}
  \bibinfo{author}{\bibnamefont{{Sheng D. N.}}}, \bibinfo{journal}{Sci. Rep.}
  \textbf{\bibinfo{volume}{4}} (\bibinfo{year}{2014}),
  \urlprefix\url{http://www.nature.com/srep/2014/140910/srep06317/abs/srep06317.html\#supplementary-information}.

\bibitem[{\citenamefont{{Zhu} et~al.}(2014)\citenamefont{{Zhu}, {Gong}, and
  {Sheng}}}]{Zhu2014}
\bibinfo{author}{\bibfnamefont{W.}~\bibnamefont{{Zhu}}},
  \bibinfo{author}{\bibfnamefont{S.~S.} \bibnamefont{{Gong}}},
  \bibnamefont{and} \bibinfo{author}{\bibfnamefont{D.~N.}
  \bibnamefont{{Sheng}}}, \bibinfo{journal}{ArXiv e-prints}
  (\bibinfo{year}{2014}), \eprint{1410.4883}.

\bibitem[{\citenamefont{{Bauer B.} et~al.}(2014)\citenamefont{{Bauer B.},
  {Cincio L.}, {Keller B.P.}, {Dolfi M.}, {Vidal G.}, {Trebst S.}, and {Ludwig
  A.W.W.}}}]{Bauer2014}
\bibinfo{author}{\bibnamefont{{Bauer B.}}},
  \bibinfo{author}{\bibnamefont{{Cincio L.}}},
  \bibinfo{author}{\bibnamefont{{Keller B.P.}}},
  \bibinfo{author}{\bibnamefont{{Dolfi M.}}},
  \bibinfo{author}{\bibnamefont{{Vidal G.}}},
  \bibinfo{author}{\bibnamefont{{Trebst S.}}}, \bibnamefont{and}
  \bibinfo{author}{\bibnamefont{{Ludwig A.W.W.}}}, \bibinfo{journal}{Nat
  Commun} \textbf{\bibinfo{volume}{5}} (\bibinfo{year}{2014}).

\bibitem[{\citenamefont{White}(1992)}]{White1992}
\bibinfo{author}{\bibfnamefont{S.~R.} \bibnamefont{White}},
  \bibinfo{journal}{Phys. Rev. Lett.} \textbf{\bibinfo{volume}{69}},
  \bibinfo{pages}{2863} (\bibinfo{year}{1992}),
  \urlprefix\url{http://link.aps.org/doi/10.1103/PhysRevLett.69.2863}.

\bibitem[{\citenamefont{Schollw\"ock}(2005)}]{Schollwock2005}
\bibinfo{author}{\bibfnamefont{U.}~\bibnamefont{Schollw\"ock}},
  \bibinfo{journal}{Rev. Mod. Phys.} \textbf{\bibinfo{volume}{77}},
  \bibinfo{pages}{259} (\bibinfo{year}{2005}),
  \urlprefix\url{http://link.aps.org/doi/10.1103/RevModPhys.77.259}.

\bibitem[{\citenamefont{Schollw{\"o}ck}(2011)}]{Schollwock2011}
\bibinfo{author}{\bibfnamefont{U.}~\bibnamefont{Schollw{\"o}ck}},
  \bibinfo{journal}{Annals of Physics} \textbf{\bibinfo{volume}{326}},
  \bibinfo{pages}{96} (\bibinfo{year}{2011}).

\bibitem[{\citenamefont{Yan et~al.}(2011)\citenamefont{Yan, Huse, and
  White}}]{Yan2011}
\bibinfo{author}{\bibfnamefont{S.}~\bibnamefont{Yan}},
  \bibinfo{author}{\bibfnamefont{D.~A.} \bibnamefont{Huse}}, \bibnamefont{and}
  \bibinfo{author}{\bibfnamefont{S.~R.} \bibnamefont{White}},
  \bibinfo{journal}{Science} \textbf{\bibinfo{volume}{332}},
  \bibinfo{pages}{1173} (\bibinfo{year}{2011}).

\bibitem[{\citenamefont{Depenbrock et~al.}(2012)\citenamefont{Depenbrock,
  McCulloch, and Schollw\"ock}}]{Depenbrock2012}
\bibinfo{author}{\bibfnamefont{S.}~\bibnamefont{Depenbrock}},
  \bibinfo{author}{\bibfnamefont{I.~P.} \bibnamefont{McCulloch}},
  \bibnamefont{and}
  \bibinfo{author}{\bibfnamefont{U.}~\bibnamefont{Schollw\"ock}},
  \bibinfo{journal}{Phys. Rev. Lett.} \textbf{\bibinfo{volume}{109}},
  \bibinfo{pages}{067201} (\bibinfo{year}{2012}),
  \urlprefix\url{http://link.aps.org/doi/10.1103/PhysRevLett.109.067201}.

\bibitem[{\citenamefont{Jiang et~al.}(2012)\citenamefont{Jiang, Wang, and
  Balents}}]{Jiang2012}
\bibinfo{author}{\bibfnamefont{H.~C.} \bibnamefont{Jiang}},
  \bibinfo{author}{\bibfnamefont{Z.}~\bibnamefont{Wang}}, \bibnamefont{and}
  \bibinfo{author}{\bibfnamefont{L.}~\bibnamefont{Balents}},
  \bibinfo{journal}{Nat. Phys.} \textbf{\bibinfo{volume}{8}},
  \bibinfo{pages}{902} (\bibinfo{year}{2012}).

\bibitem[{\citenamefont{Freedman et~al.}(2004)\citenamefont{Freedman, Nayak,
  Shtengel, Walker, and Wang}}]{Freedman2004}
\bibinfo{author}{\bibfnamefont{M.}~\bibnamefont{Freedman}},
  \bibinfo{author}{\bibfnamefont{C.}~\bibnamefont{Nayak}},
  \bibinfo{author}{\bibfnamefont{K.}~\bibnamefont{Shtengel}},
  \bibinfo{author}{\bibfnamefont{K.}~\bibnamefont{Walker}}, \bibnamefont{and}
  \bibinfo{author}{\bibfnamefont{Z.}~\bibnamefont{Wang}},
  \bibinfo{journal}{Annals of Physics} \textbf{\bibinfo{volume}{310}},
  \bibinfo{pages}{428 } (\bibinfo{year}{2004}), ISSN \bibinfo{issn}{0003-4916},
  \urlprefix\url{http://www.sciencedirect.com/science/article/pii/S0003491604000260}.

\bibitem[{\citenamefont{Levin and Wen}(2005)}]{Levin2005}
\bibinfo{author}{\bibfnamefont{M.~A.} \bibnamefont{Levin}} \bibnamefont{and}
  \bibinfo{author}{\bibfnamefont{X.-G.} \bibnamefont{Wen}},
  \bibinfo{journal}{Phys. Rev. B} \textbf{\bibinfo{volume}{71}},
  \bibinfo{pages}{045110} (\bibinfo{year}{2005}),
  \urlprefix\url{http://link.aps.org/doi/10.1103/PhysRevB.71.045110}.

\bibitem[{\citenamefont{Poilblanc et~al.}(2012)\citenamefont{Poilblanc, Schuch,
  Perez-Garcia, and Cirac}}]{Poilblanc2012}
\bibinfo{author}{\bibfnamefont{D.}~\bibnamefont{Poilblanc}},
  \bibinfo{author}{\bibfnamefont{N.}~\bibnamefont{Schuch}},
  \bibinfo{author}{\bibfnamefont{D.}~\bibnamefont{Perez-Garcia}},
  \bibnamefont{and} \bibinfo{author}{\bibfnamefont{J.~I.} \bibnamefont{Cirac}},
  \bibinfo{journal}{Phys. Rev. B} \textbf{\bibinfo{volume}{86}},
  \bibinfo{pages}{014404} (\bibinfo{year}{2012}),
  \urlprefix\url{http://link.aps.org/doi/10.1103/PhysRevB.86.014404}.

\bibitem[{\citenamefont{Schuch et~al.}(2012)\citenamefont{Schuch, Poilblanc,
  Cirac, and Perez-Garcia}}]{Schuch2012}
\bibinfo{author}{\bibfnamefont{N.}~\bibnamefont{Schuch}},
  \bibinfo{author}{\bibfnamefont{D.}~\bibnamefont{Poilblanc}},
  \bibinfo{author}{\bibfnamefont{J.~I.} \bibnamefont{Cirac}}, \bibnamefont{and}
  \bibinfo{author}{\bibfnamefont{D.}~\bibnamefont{Perez-Garcia}},
  \bibinfo{journal}{Phys. Rev. B} \textbf{\bibinfo{volume}{86}},
  \bibinfo{pages}{115108} (\bibinfo{year}{2012}),
  \urlprefix\url{http://link.aps.org/doi/10.1103/PhysRevB.86.115108}.

\bibitem[{\citenamefont{{He} and {Chen}}(2014)}]{He2014a}
\bibinfo{author}{\bibfnamefont{Y.-C.} \bibnamefont{{He}}} \bibnamefont{and}
  \bibinfo{author}{\bibfnamefont{Y.}~\bibnamefont{{Chen}}},
  \bibinfo{journal}{ArXiv e-prints}  (\bibinfo{year}{2014}),
  \eprint{1407.2740}.

\bibitem[{\citenamefont{{Qi} et~al.}(2014)\citenamefont{{Qi}, {Gu}, and
  {Yao}}}]{Qi2014}
\bibinfo{author}{\bibfnamefont{Y.}~\bibnamefont{{Qi}}},
  \bibinfo{author}{\bibfnamefont{Z.-C.} \bibnamefont{{Gu}}}, \bibnamefont{and}
  \bibinfo{author}{\bibfnamefont{H.}~\bibnamefont{{Yao}}},
  \bibinfo{journal}{ArXiv e-prints}  (\bibinfo{year}{2014}),
  \eprint{1406.6364}.

\bibitem[{\citenamefont{{Buerschaper} et~al.}(2014)\citenamefont{{Buerschaper},
  {Morampudi}, and {Pollmann}}}]{Buerschaper2014}
\bibinfo{author}{\bibfnamefont{O.}~\bibnamefont{{Buerschaper}}},
  \bibinfo{author}{\bibfnamefont{S.~C.} \bibnamefont{{Morampudi}}},
  \bibnamefont{and}
  \bibinfo{author}{\bibfnamefont{F.}~\bibnamefont{{Pollmann}}},
  \bibinfo{journal}{ArXiv e-prints}  (\bibinfo{year}{2014}),
  \eprint{1407.8521}.

\bibitem[{\citenamefont{{Iqbal} et~al.}(2014)\citenamefont{{Iqbal},
  {Poilblanc}, and {Schuch}}}]{Iqbal_M2014}
\bibinfo{author}{\bibfnamefont{M.}~\bibnamefont{{Iqbal}}},
  \bibinfo{author}{\bibfnamefont{D.}~\bibnamefont{{Poilblanc}}},
  \bibnamefont{and} \bibinfo{author}{\bibfnamefont{N.}~\bibnamefont{{Schuch}}},
  \bibinfo{journal}{ArXiv e-prints}  (\bibinfo{year}{2014}),
  \eprint{1407.7773}.

\bibitem[{\citenamefont{{Zaletel} and {Vishwanath}}(2014)}]{Zalatel2014}
\bibinfo{author}{\bibfnamefont{M.~P.} \bibnamefont{{Zaletel}}}
  \bibnamefont{and}
  \bibinfo{author}{\bibfnamefont{A.}~\bibnamefont{{Vishwanath}}},
  \bibinfo{journal}{ArXiv e-prints}  (\bibinfo{year}{2014}),
  \eprint{1410.2894}.

\bibitem[{\citenamefont{Chalker et~al.}(1992)\citenamefont{Chalker, Holdsworth,
  and Shender}}]{Chalker1992}
\bibinfo{author}{\bibfnamefont{J.~T.} \bibnamefont{Chalker}},
  \bibinfo{author}{\bibfnamefont{P.~C.~W.} \bibnamefont{Holdsworth}},
  \bibnamefont{and} \bibinfo{author}{\bibfnamefont{E.~F.}
  \bibnamefont{Shender}}, \bibinfo{journal}{Phys. Rev. Lett.}
  \textbf{\bibinfo{volume}{68}}, \bibinfo{pages}{855} (\bibinfo{year}{1992}),
  \urlprefix\url{http://link.aps.org/doi/10.1103/PhysRevLett.68.855}.

\bibitem[{\citenamefont{Huse and Rutenberg}(1992)}]{Huse1992}
\bibinfo{author}{\bibfnamefont{D.~A.} \bibnamefont{Huse}} \bibnamefont{and}
  \bibinfo{author}{\bibfnamefont{A.~D.} \bibnamefont{Rutenberg}},
  \bibinfo{journal}{Phys. Rev. B} \textbf{\bibinfo{volume}{45}},
  \bibinfo{pages}{7536} (\bibinfo{year}{1992}),
  \urlprefix\url{http://link.aps.org/doi/10.1103/PhysRevB.45.7536}.

\bibitem[{\citenamefont{Ritchey et~al.}(1993)\citenamefont{Ritchey, Chandra,
  and Coleman}}]{Ritchey1993}
\bibinfo{author}{\bibfnamefont{I.}~\bibnamefont{Ritchey}},
  \bibinfo{author}{\bibfnamefont{P.}~\bibnamefont{Chandra}}, \bibnamefont{and}
  \bibinfo{author}{\bibfnamefont{P.}~\bibnamefont{Coleman}},
  \bibinfo{journal}{Phys. Rev. B} \textbf{\bibinfo{volume}{47}},
  \bibinfo{pages}{15342} (\bibinfo{year}{1993}),
  \urlprefix\url{http://link.aps.org/doi/10.1103/PhysRevB.47.15342}.

\bibitem[{\citenamefont{Messio et~al.}(2011)\citenamefont{Messio, Lhuillier,
  and Misguich}}]{Messio2011}
\bibinfo{author}{\bibfnamefont{L.}~\bibnamefont{Messio}},
  \bibinfo{author}{\bibfnamefont{C.}~\bibnamefont{Lhuillier}},
  \bibnamefont{and} \bibinfo{author}{\bibfnamefont{G.}~\bibnamefont{Misguich}},
  \bibinfo{journal}{Phys. Rev. B} \textbf{\bibinfo{volume}{83}},
  \bibinfo{pages}{184401} (\bibinfo{year}{2011}),
  \urlprefix\url{http://link.aps.org/doi/10.1103/PhysRevB.83.184401}.

\bibitem[{\citenamefont{Cepas and Ralko}(2011)}]{Cepas2011}
\bibinfo{author}{\bibfnamefont{O.}~\bibnamefont{Cepas}} \bibnamefont{and}
  \bibinfo{author}{\bibfnamefont{A.}~\bibnamefont{Ralko}},
  \bibinfo{journal}{Phys. Rev. B} \textbf{\bibinfo{volume}{84}},
  \bibinfo{pages}{020413} (\bibinfo{year}{2011}),
  \urlprefix\url{http://link.aps.org/doi/10.1103/PhysRevB.84.020413}.

\bibitem[{\citenamefont{Spenke and Guertler}(2012)}]{Spenke2012}
\bibinfo{author}{\bibfnamefont{M.}~\bibnamefont{Spenke}} \bibnamefont{and}
  \bibinfo{author}{\bibfnamefont{S.}~\bibnamefont{Guertler}},
  \bibinfo{journal}{Phys. Rev. B} \textbf{\bibinfo{volume}{86}},
  \bibinfo{pages}{054440} (\bibinfo{year}{2012}),
  \urlprefix\url{http://link.aps.org/doi/10.1103/PhysRevB.86.054440}.

\bibitem[{\citenamefont{Chern and Moessner}(2013)}]{Chern2013}
\bibinfo{author}{\bibfnamefont{G.-W.} \bibnamefont{Chern}} \bibnamefont{and}
  \bibinfo{author}{\bibfnamefont{R.}~\bibnamefont{Moessner}},
  \bibinfo{journal}{Phys. Rev. Lett.} \textbf{\bibinfo{volume}{110}},
  \bibinfo{pages}{077201} (\bibinfo{year}{2013}),
  \urlprefix\url{http://link.aps.org/doi/10.1103/PhysRevLett.110.077201}.

\bibitem[{\citenamefont{Harris et~al.}(1992)\citenamefont{Harris, Kallin, and
  Berlinsky}}]{Harris1992}
\bibinfo{author}{\bibfnamefont{A.~B.} \bibnamefont{Harris}},
  \bibinfo{author}{\bibfnamefont{C.}~\bibnamefont{Kallin}}, \bibnamefont{and}
  \bibinfo{author}{\bibfnamefont{A.~J.} \bibnamefont{Berlinsky}},
  \bibinfo{journal}{Phys. Rev. B} \textbf{\bibinfo{volume}{45}},
  \bibinfo{pages}{2899} (\bibinfo{year}{1992}), \urlprefix\url{Harris}.

\bibitem[{\citenamefont{Lecheminant et~al.}(1997)\citenamefont{Lecheminant,
  Bernu, Lhuillier, Pierre, and Sindzingre}}]{Lecheminant1997}
\bibinfo{author}{\bibfnamefont{P.}~\bibnamefont{Lecheminant}},
  \bibinfo{author}{\bibfnamefont{B.}~\bibnamefont{Bernu}},
  \bibinfo{author}{\bibfnamefont{C.}~\bibnamefont{Lhuillier}},
  \bibinfo{author}{\bibfnamefont{L.}~\bibnamefont{Pierre}}, \bibnamefont{and}
  \bibinfo{author}{\bibfnamefont{P.}~\bibnamefont{Sindzingre}},
  \bibinfo{journal}{Phys. Rev. B} \textbf{\bibinfo{volume}{56}},
  \bibinfo{pages}{2521} (\bibinfo{year}{1997}),
  \urlprefix\url{http://link.aps.org/doi/10.1103/PhysRevB.56.2521}.

\bibitem[{\citenamefont{Suttner et~al.}(2014)\citenamefont{Suttner, Platt,
  Reuther, and Thomale}}]{Suttner2014}
\bibinfo{author}{\bibfnamefont{R.}~\bibnamefont{Suttner}},
  \bibinfo{author}{\bibfnamefont{C.}~\bibnamefont{Platt}},
  \bibinfo{author}{\bibfnamefont{J.}~\bibnamefont{Reuther}}, \bibnamefont{and}
  \bibinfo{author}{\bibfnamefont{R.}~\bibnamefont{Thomale}},
  \bibinfo{journal}{Phys. Rev. B} \textbf{\bibinfo{volume}{89}},
  \bibinfo{pages}{020408} (\bibinfo{year}{2014}),
  \urlprefix\url{http://link.aps.org/doi/10.1103/PhysRevB.89.020408}.

\bibitem[{\citenamefont{Li and Haldane}(2008)}]{Li2008}
\bibinfo{author}{\bibfnamefont{H.}~\bibnamefont{Li}} \bibnamefont{and}
  \bibinfo{author}{\bibfnamefont{F.~D.~M.} \bibnamefont{Haldane}},
  \bibinfo{journal}{Phys. Rev. Lett.} \textbf{\bibinfo{volume}{101}},
  \bibinfo{pages}{010504} (\bibinfo{year}{2008}),
  \urlprefix\url{http://link.aps.org/doi/10.1103/PhysRevLett.101.010504}.

\bibitem[{\citenamefont{Metlitski and Grover}(2011)}]{Metlitski2011}
\bibinfo{author}{\bibfnamefont{M.~A.} \bibnamefont{Metlitski}}
  \bibnamefont{and} \bibinfo{author}{\bibfnamefont{T.}~\bibnamefont{Grover}},
  \bibinfo{journal}{arXiv:1112.5166}  (\bibinfo{year}{2011}).

\bibitem[{\citenamefont{Alba et~al.}(2013)\citenamefont{Alba, Haque, and
  L\"auchli}}]{Alba2013}
\bibinfo{author}{\bibfnamefont{V.}~\bibnamefont{Alba}},
  \bibinfo{author}{\bibfnamefont{M.}~\bibnamefont{Haque}}, \bibnamefont{and}
  \bibinfo{author}{\bibfnamefont{A.~M.} \bibnamefont{L\"auchli}},
  \bibinfo{journal}{Phys. Rev. Lett.} \textbf{\bibinfo{volume}{110}},
  \bibinfo{pages}{260403} (\bibinfo{year}{2013}),
  \urlprefix\url{http://link.aps.org/doi/10.1103/PhysRevLett.110.260403}.

\bibitem[{\citenamefont{Kolley et~al.}(2013)\citenamefont{Kolley, Depenbrock,
  McCulloch, Schollw\"ock, and Alba}}]{Kolley2013}
\bibinfo{author}{\bibfnamefont{F.}~\bibnamefont{Kolley}},
  \bibinfo{author}{\bibfnamefont{S.}~\bibnamefont{Depenbrock}},
  \bibinfo{author}{\bibfnamefont{I.~P.} \bibnamefont{McCulloch}},
  \bibinfo{author}{\bibfnamefont{U.}~\bibnamefont{Schollw\"ock}},
  \bibnamefont{and} \bibinfo{author}{\bibfnamefont{V.}~\bibnamefont{Alba}},
  \bibinfo{journal}{Phys. Rev. B} \textbf{\bibinfo{volume}{88}},
  \bibinfo{pages}{144426} (\bibinfo{year}{2013}),
  \urlprefix\url{http://link.aps.org/doi/10.1103/PhysRevB.88.144426}.

\bibitem[{\citenamefont{Sandvik}(2010)}]{Sandvik2010}
\bibinfo{author}{\bibfnamefont{A.~W.} \bibnamefont{Sandvik}},
  \bibinfo{journal}{AIP Conference Proceedings}
  \textbf{\bibinfo{volume}{1297}}, \bibinfo{pages}{135} (\bibinfo{year}{2010}),
  \urlprefix\url{http://scitation.aip.org/content/aip/proceeding/aipcp/10.1063/1.3518900}.

\bibitem[{\citenamefont{Pelissetto and Vicari}(2002)}]{Pelissetto2002}
\bibinfo{author}{\bibfnamefont{A.}~\bibnamefont{Pelissetto}} \bibnamefont{and}
  \bibinfo{author}{\bibfnamefont{E.}~\bibnamefont{Vicari}},
  \bibinfo{journal}{Physics Reports} \textbf{\bibinfo{volume}{368}},
  \bibinfo{pages}{549 } (\bibinfo{year}{2002}), ISSN \bibinfo{issn}{0370-1573},
  \urlprefix\url{http://www.sciencedirect.com/science/article/pii/S0370157302002193}.

\bibitem[{\citenamefont{Sandvik}(2012)}]{Sandvik2012}
\bibinfo{author}{\bibfnamefont{A.~W.} \bibnamefont{Sandvik}},
  \bibinfo{journal}{Phys. Rev. B} \textbf{\bibinfo{volume}{85}},
  \bibinfo{pages}{134407} (\bibinfo{year}{2012}),
  \urlprefix\url{http://link.aps.org/doi/10.1103/PhysRevB.85.134407}.

\bibitem[{\citenamefont{{Gong} et~al.}(2014)\citenamefont{{Gong}, {Zhu},
  {Balents}, and {Sheng}}}]{Gong2014b}
\bibinfo{author}{\bibfnamefont{S.-S.} \bibnamefont{{Gong}}},
  \bibinfo{author}{\bibfnamefont{W.}~\bibnamefont{{Zhu}}},
  \bibinfo{author}{\bibfnamefont{L.}~\bibnamefont{{Balents}}},
  \bibnamefont{and} \bibinfo{author}{\bibfnamefont{D.~N.}
  \bibnamefont{{Sheng}}}, \bibinfo{journal}{ArXiv e-prints}
  (\bibinfo{year}{2014}), \eprint{1412.1571}.

\bibitem[{\citenamefont{Iqbal et~al.}(2015)\citenamefont{Iqbal, Poilblanc, and
  Becca}}]{Iqbal2015}
\bibinfo{author}{\bibfnamefont{Y.}~\bibnamefont{Iqbal}},
  \bibinfo{author}{\bibfnamefont{D.}~\bibnamefont{Poilblanc}},
  \bibnamefont{and} \bibinfo{author}{\bibfnamefont{F.}~\bibnamefont{Becca}},
  \bibinfo{journal}{Phys. Rev. B} \textbf{\bibinfo{volume}{91}},
  \bibinfo{pages}{020402} (\bibinfo{year}{2015}),
  \urlprefix\url{http://link.aps.org/doi/10.1103/PhysRevB.91.020402}.

\end{thebibliography}

\end{document}